\documentclass[12pt]{iopart}

\usepackage{hyperref}
\usepackage{graphicx}
\usepackage{dcolumn}
\usepackage{bm}
\usepackage{xcolor} 
\usepackage{textcomp} 
\usepackage[export]{adjustbox} 
\usepackage[english]{babel}

\usepackage{hhline}
\usepackage{comment}
\usepackage{color}
\usepackage{amsfonts}
\usepackage{amssymb}
\usepackage{amsmath}

\usepackage{dsfont}
\usepackage{graphicx}
\usepackage{multirow}
\usepackage{pbox}

\usepackage[mathlines]{lineno}

\let\oldequation\equation
\let\oldendequation\endequation
\let\oldalign\align  
\let\oldendalign\endalign  

\renewenvironment{equation}
  {\linenomathNonumbers\oldequation}
  {\oldendequation\endlinenomath}
\renewenvironment{align}
  {\linenomathNonumbers\oldalign}
  {\oldendalign\endlinenomath}

\newcommand{\kin}{{\rm kin}}

\usepackage{xr}
\begin{document}
\title[Table-top 3D photoemission orbital tomography with a fs-EUV light source]{Table-top three-dimensional photoemission orbital tomography with a femtosecond extreme ultraviolet light source}

\author{
W~Bennecke$^1$, 
TL~Dinh$^2$,
J~P~Bange$^1$, D~Schmitt$^1$, M~Merboldt$^1$, L~Weinhagen$^1$,  B~van~Wingerden$^1$,
F~Frassetto$^3$, L~Poletto$^3$,
M~Reutzel$^1$, D~Steil$^1$, 
D~R~Luke$^2$, 
S~Mathias$^{1,4,*}$ and 
G~S~M~Jansen$^{1,*}$}

\address{$^1$ 1\textsuperscript{st} Institute of Physics, University of G\"ottingen, Friedrich-Hund-Platz 1, 37077, G\"ottingen, Germany}
\address{$^2$ Institute for Numerical and Applied Mathematics, University of G\"ottingen, Lotzestrasse 16-18, 37083 G\"ottingen , Germany}
\address{$^3$ Institute for Photonics and Nanotechnologies CNR-IFN, 35131 Padova, Italy}
\address{$^4$ International Center for Advanced Studies of Energy Conversion (ICASEC), University of Göttingen, Göttingen, Germany}
\ead{smathias@uni-goettingen.de; gsmjansen@uni-goettingen.de}










\begin{abstract}
Following electronic processes in molecules and materials at the level of the quantum mechanical electron wavefunction with \aa{}ngstr\"om-level spatial resolution and with full access to its femtosecond temporal dynamics is at the heart of ultrafast condensed matter physics. A breakthrough invention allowing experimental access to electron wavefunctions was the reconstruction of molecular orbitals from angle-resolved photoelectron spectroscopy data in 2009, termed photoemission orbital tomography (POT). This invention puts ultrafast three-dimensional (3D) POT in reach, with many new prospects for the study of ultrafast light-matter interaction, femtochemistry and photo-induced phase transitions. 
Here, we develop a synergistic experimental-algorithmic approach to realize the first 3D-POT experiment using a short-pulse extreme ultraviolet light source. We combine a new variant of photoelectron spectroscopy, namely ultrafast momentum microscopy, with a table-top spectrally-tunable high-harmonic generation light source and a tailored algorithm for efficient 3D reconstruction from sparse, undersampled data. This combination dramatically speeds up the experimental data acquisition, while at the same time reducing the sampling requirements to achieve complete 3D information. We demonstrate the power of this approach by full 3D imaging of the frontier orbitals of a prototypical organic semiconductor absorbed on pristine Ag(110). 

\end{abstract}

\maketitle

\section{Introduction}
\begin{figure}[bth]
    \centering
    \vspace{.4cm}
    \includegraphics[width=.66\textwidth]{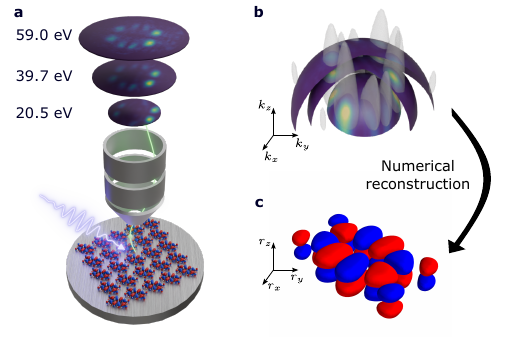}
    \caption{\textbf{Schematic overview of a three-dimensional photoemission orbital tomography experiment. }
    a) A photon with energy $h\nu$ photoexcites an electron out of the highest occupied molecular orbital of a well-ordered molecular layer, and a photoemission momentum microscope is used to record the full momentum- and energy-dependent photoelectron spectrum. Repeating this measurement for several $h\nu$ yields a 3D fingerprint of the molecular orbital.
    b) According to the plane-wave model of photoemission, each photoelectron kinetic energy yields the intensity distribution on a hemispherical shell in the 3D momentum space of the molecular orbital. c) Numerical reconstruction of the three-dimensional HOMO of PTCDA from the measured sparse data recorded at only seven photon energies.}
    \label{fig:schematic_3D-POT}
\end{figure}

The quantum mechanical electron wavefunction determines fundamental properties of matter such as electronic structure, chemical bonding and light-matter interaction. In consequence, an in-depth characterization of its properties offers unique insight into electronic and optoelectronic functionalities of materials. Photoemission orbital tomography (POT) has appeared as a uniquely sensitive probe for the electronic wavefunction in organic molecular systems \cite{puschnig_reconstruction_2009}. Nowadays, POT allows studying optically excited~\cite{wallauer_tracing_2021, adamkiewicz_coherent_2023, baumgartner_ultrafast_2022} and excitonic states~\cite{neef_orbital-resolved_2022, bennecke_multiorbital_2023, schmitt_formation_2022, madeo_directly_2020, dong_direct_2021, bennecke_hybrid_2024}, characterizing molecular structure, absorption geometry and interlayer hybridization in hybrid interfaces~\cite{yang_identifying_2019, hurdax_large_2022, yang_momentum-selective_2022} and the direct imaging of molecular orbitals independent of theoretical calculations \cite{puschnig_reconstruction_2009, luftner_imaging_2014, kliuiev_application_2016, jansen_efficient_2020}. Amongst a range of techniques that provide access to the real-space distribution of the electron orbitals~\cite{repp_molecules_2005, cocker_tracking_2016}, POT and the related gas-phase molecular orbital tomography~\cite{itatani_tomographic_2004, vozzi_generalized_2011} stand furthermore out due to a unique capability: they provide non-invasive access to the intrinsic three-dimensional shape of the molecular orbital \cite{weis_exploring_2015, graus_three-dimensional_2019}. The mechanism by which three-dimensional information is acquired is the same for both methods: full three-dimensional electron momentum distributions are acquired by recording photoelectron spectra for a range of photon energies. As sketched in Fig.~\ref{fig:schematic_3D-POT}, analysis of these data enables to image the full three-dimensional (3D) molecular orbital in real space with sub-\aa{}ngström resolution. 

However, while all of these studies have yielded ground-breaking results, a long-awaited extension of POT would combine full 3D orbital reconstruction with a tunable femtosecond extreme ultraviolet (EUV) light source that would not only enable three-dimensional orbital imaging, but also time-resolved studies at the laboratory scale. Such an experiment has many potential applications in different research areas of (ultrafast) condensed matter physics, for example for the study of orbital hybridization in organic-inorganic heterostructures~\cite{yang_momentum-selective_2022, krumland_ab_initio_2024}, for time-resolved studies of exciton dynamics in organic semiconductors~\cite{neef_orbital-resolved_2022, bennecke_multiorbital_2023}, and for motion-induced orbital dynamics~\cite{cocker_tracking_2016, baumgartner_multiplex_2023}. 

For this reason, first 3D photoemission orbital tomography (3D-POT) studies have been undertaken \cite{hurdax_large_2022} and successfully imaged the two highest occupied molecular orbitals of PTCDA (full name: perylene-3,4,9,10-tetracarboxylic dianhydride) adsorbed on the Ag(110) surface \cite{weis_exploring_2015, graus_three-dimensional_2019}. However, the nature of 3D-POT measurements, requiring photon-energy-dependent data collection in the EUV range, has so far prevented wide-spread adoption and restricted the experiments to large-scale synchrotron facilities. Moreover, in order to measure complete and high-resolution momentum distributions for a given orbital, 3D-POT seemingly relies on recording two-dimensional photoemission momentum maps for a large number of photon energies~\cite{weis_exploring_2015, graus_three-dimensional_2019}. While this tedious data acquisition was already sped up through the use of momentum microscopy \cite{jansen_efficient_2020, graus_three-dimensional_2019}, the vast amount of data needed to reconstruct 3D information and, for time-resolved experiments, the need to scan several pump-probe delays has so far strongly limited the potential of 3D-POT.

In this article, we overcome all of these limitations and demonstrate how 3D-POT can be performed in the laboratory at measurement durations that even permit the application of ultrafast 3D orbital imaging. To achieve this, we have combined time-of-flight (ToF) photoemission momentum microscopy~\cite{keunecke_time-resolved_2020} with a photon-energy-tunable femtosecond high-harmonic generation light source covering the photon energy range from 13 to 71~eV~\cite{frassetto_single-grating_2011}. This ToF-based setup, which does not rely on scanning the energy or momentum of the photoelectrons, allows us to collect the photon-energy-dependent momentum distributions needed for 3D reconstruction much more efficiently. Importantly, we complement this experimental strategy with an advanced numerical reconstruction algorithm that requires much less photon energies for full 3D orbital imaging~\cite{dinh_minimalist_2024}. All in all, this joint experimental-algorithmic development establishes lab-scale streamlined 3D-POT, which we benchmark by 3D imaging of the frontier molecular orbitals of PTCDA/Ag(110). Our results clearly confirm the potential of table-top 3D-POT for systematic studies of orbital hybridization and even for full, femtosecond time-resolved orbital tomography.

\section{Light source for an ultrafast 3D-POT experiment}
The requirements for the light source in a 3D-POT experiment are directly determined from the desired resolution of the reconstructed molecular orbital in 3D-POT. According to the commonly used plane-wave approximation of the photoemission process \cite{puschnig_reconstruction_2009, gadzuk_surface_1974}, the measured photoelectron current $I_{h\nu}$ in a POT experiment can be expressed as
\begin{equation}
    I_{h\nu}(\vec{k}) = |\vec{A}\cdot \vec{k}|^2~ |\mathcal{F}(\psi)(\vec{k})|^2~\delta(E_{\rm I} + E_{\kin} - h\nu)\,.
    \label{eq:plane_wave_model}
\end{equation}
This equation highlights the well-known relation that the momentum-resolved photoelectron distribution is proportional to the Fourier transform $\mathcal{F}$ of the real-space molecular orbital $\psi$, modulated by a weakly varying polarization factor $|\vec{A}\cdot \vec{k}|$.
Here, $\vec{A}$ is the vector potential of the incident (ionizing) light, which has photon energy $h\nu$, and $\vec{k}$ is the photoelectron momentum. 
The Dirac delta in Eq.~\ref{eq:plane_wave_model} expresses the energy conservation in terms of the photon energy $h\nu$, the orbital ionization energy $E_{\rm I}$ and the photoelectron kinetic energy $E_{\kin}$. As the latter is proportional to $\hbar^2 k^2$, it follows that an angle-resolved photoelectron spectroscopy (ARPES) measurement at a fixed photon energy measures the value of $|\mathcal{F}(\psi)(\vec{k})|$ on a hemisphere in momentum space (see Fig.~\ref{fig:schematic_3D-POT}b). In the absence of an inner potential, the radius of this hemisphere can be approximated by
\begin{linenomath*}
\begin{equation*}
    |k({\rm \AA^{-1}})| \approx 0.514\sqrt{E_\kin({\rm eV})}.
\end{equation*}
\end{linenomath*}
In POT, this relation implies that photon energies of at least 15-20~eV are necessary to probe the features of valence orbitals in $\pi$-conjugated organic molecules, as their brightest photoemission signatures usually appear around $\approx2$~\AA\textsuperscript{-1}. Similarly, the resolution in 3D (and 2D) orbital imaging can be directly estimated from this relation, and it follows that the photon energy must be approximately quadrupled to achieve a doubling of the final image resolution. It can thus be concluded that \aa{}ngström-scale 3D-POT requires a broad range of extreme ultraviolet (EUV) photon energies, starting at 15~eV and extending upwards of 60~eV. These photon energies can be generated efficiently using laser-based high-harmonic generation (HHG) \cite{krausz_attosecond_2009, mathias_angle-resolved_2007}, which conveniently provides a short-pulse table-top solution that enables time-resolved spectroscopies, too~\cite{wallauer_tracing_2021, keunecke_time-resolved_2020, maklar_quantitative_2020, kunin_momentum-resolved_2023, reutzel_probing_2024, karni_through_2023, heber_multispectral_2022}.

To generate HHG over the desired spectral window, we tightly focus 65~W output of our Yb-fiber laser amplifier (35~fs pulses, 500~kHz repetition rate, 130~\textmu J pulse energy) using an $f=7.5$~cm lens into an argon gas jet, which has sufficiently high ionization potential to achieve a $>71$~eV cutoff while also affording good HHG efficiency. 
In order to isolate individual high harmonics and to apply these in the photoelectron momentum microscope, we implemented a grating-based EUV monochromator in grazing-incidence, off-plane diffraction geometry~\cite{frassetto_single-grating_2011, kunin_momentum-resolved_2023, grazioli_CITIUS_2014} (see Supplemental Information). 
In this way, a high photon efficiency can be achieved for both $s$ and $p$ polarization, while the grating line spacing can be minimized efficiently to reduce the effects of geometric temporal broadening of the femtosecond EUV pulses~\cite{poletto_ultrafast_2012}. 
The design of the single-grating monochromator used in this study is shown in Supplementary Fig.~\ref{fig:momichromator_setup} and described in detail in the Methods section \ref{sec:MoMichromator}. In short, a selection of four different plane gratings (100, 300, 400, 600 lines/mm, respectively) allows us to choose any harmonic from our HHG light source between 13~eV to 71~eV with minimized temporal broadening. Thus, we can perform time-resolved momentum microscopy using 25 different photon energies with a spacing of 2.4 eV within this energy range.

\section{Table-top 3D-POT data collection}
We demonstrate the 3D-POT capabilities for the prototypical hybrid interface of PTCDA adsorbed on the Ag(110) surface~\cite{luftner_imaging_2014, kliuiev_application_2016, weis_exploring_2015, graus_three-dimensional_2019}. For this interface, the first PTCDA monolayer adsorbs in a highly ordered brick-wall structure, where furthermore the lowest unoccupied molecular orbital (LUMO) of the gas-phase molecule is occupied due to charge transfer from the substrate~\cite{wiesner_electronic_2012}. Thus, the ToF momentum microscope allows to measure orbital fingerprints of both the highest occupied molecular orbital (HOMO) and the LUMO simultaneously~\cite{jansen_efficient_2020}. 

\begin{figure}[htb]
    \centering
    \includegraphics[width=\linewidth]{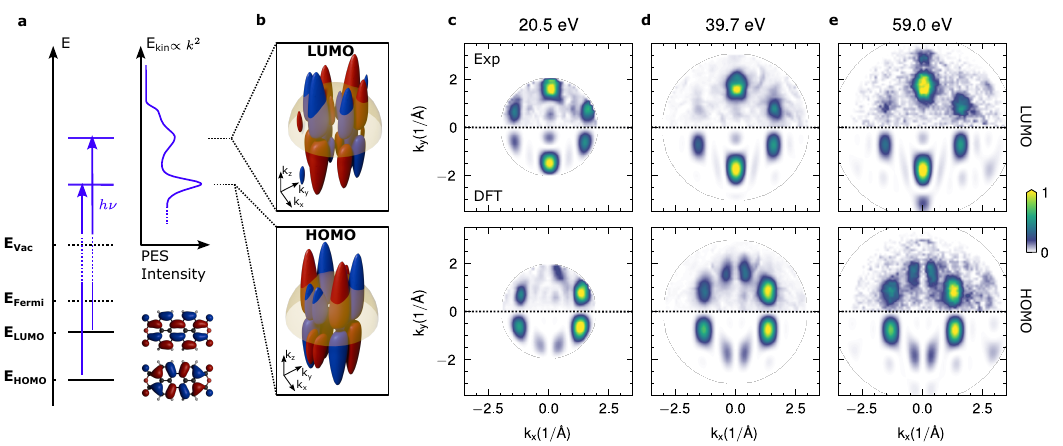}  
    \caption{\textbf{Exemplary 3D-POT data of the frontier orbitals of PTCDA adsorbed on Ag(110). }
    (a) Energy alignment of the the vacuum and the Fermi level and the HOMO and LUMO states, indicating that the LUMO shifts below the Fermi level and is occupied due to charge transfer from the silver surface. Electrons from these states are photo-excited using EUV light with photon energy $h\nu$, giving rise to a photoelectron spectrum such as sketched in the top right. The bottom right of (a) shows calculated isosurfaces of the HOMO (bottom) and LUMO (top) wavefunctions of PTCDA.
    (b) For each photon energy, we record the intensity on a hemisphere (yellow shaded half-sphere) in 3D momentum space, such that the radius of the hemisphere $|k|$ is given by $\sqrt{h\nu - E_{\rm HOMO/LUMO}}$. 
    (c-e) Typical photoemission momentum maps recorded at photon energies $h\nu=20.5$ (c), $39.7$ (d) and $59.0$~eV (e) for the HOMO (bottom row) and LUMO (top row). In each momentum map, the top half ($k_y>0$) shows the experimental data, while the bottom half ($k_y<0$) shows the predicted momentum fingerprint for a gas-phase PTCDA molecule from density functional theory. Each momentum fingerprint has been normalized to the maximal intensity feature for visualization only.}
    \label{fig:3D-POT_data}
\end{figure}

In order to have a comprehensive data set to test our setup and the developed 3D-POT algorithms, we measured the full momentum-energy-resolved photoelectron spectrum of PTCDA/Ag(110) for ten different EUV photon energies ranging from 20.5~eV to 63.8~eV (cf. Methods, Table~\ref{tab:Photonenergies}). Each spectrum was recorded within two hours at a typical count rate of $2\times 10^5$ photoelectrons per second (integrated over the full spectrum, laser repetition rate of 500 kHz). 
We reiterate that the ToF-based momentum microscope records the full 2D momentum and kinetic energy of each individual electron that is accepted into the microscope. In practice, a single data set contains 3D ($k_x,k_y,E_{\rm kin}$) data in an energy range of about 5~eV and over a circular momentum region with a radius of about 3~\AA{}\textsuperscript{-1}, so that no additional scanning in kinetic energy or momentum is necessary. In addition, we emphasize that we measured ten photon-energy-dependent data sets to be able to thoroughly test our algorithms, but that one of the most important and surprising results of our work is that, by using our iterative reconstruction algorithm, even just four data sets can be sufficient for a full 3D reconstruction of the orbitals.

Before we can start the real-space reconstruction, we first need to determine the energy of each orbital in the photoelectron spectrum (Fig.~\ref{fig:3D-POT_data}a) and extract the corresponding 2D momentum fingerprint for each used photon energy, which is then mapped to hemispherical shells as indicated in Fig.~\ref{fig:3D-POT_data}b. 
To do so, we model the photoelectron spectrum at each momentum ($k_x,k_y$) by a Gaussian lineshape of the molecular orbital fingerprint overlapping with a smooth Fermi-Dirac distribution, and fit the relative amplitudes for each position. A full description of the fit routine to extract the orbital momentum fingerprint, i.e. to remove all background other than the intensity of the orbital itself, is given in the Methods section~\ref{sec:methods:background}. Fig.~\ref{fig:3D-POT_data}c-e shows the background-corrected momentum maps of the LUMO and HOMO for three exemplary photon energies. The experimental data (upper half of each panel) is compared to simulated momentum maps of the gas phase PTCDA molecule based on Kohn-Sham (KS) orbitals retrieved from Ref.~\cite{puschnig_molecular_database} (lower half of each panel), and is in excellent agreement. After normalization of the extracted momentum maps to the measured photon flux, the data is ready for iterative reconstruction of the real-space molecular orbitals. 

\begin{figure}[htb]
    \centering
     \includegraphics[width=\linewidth]{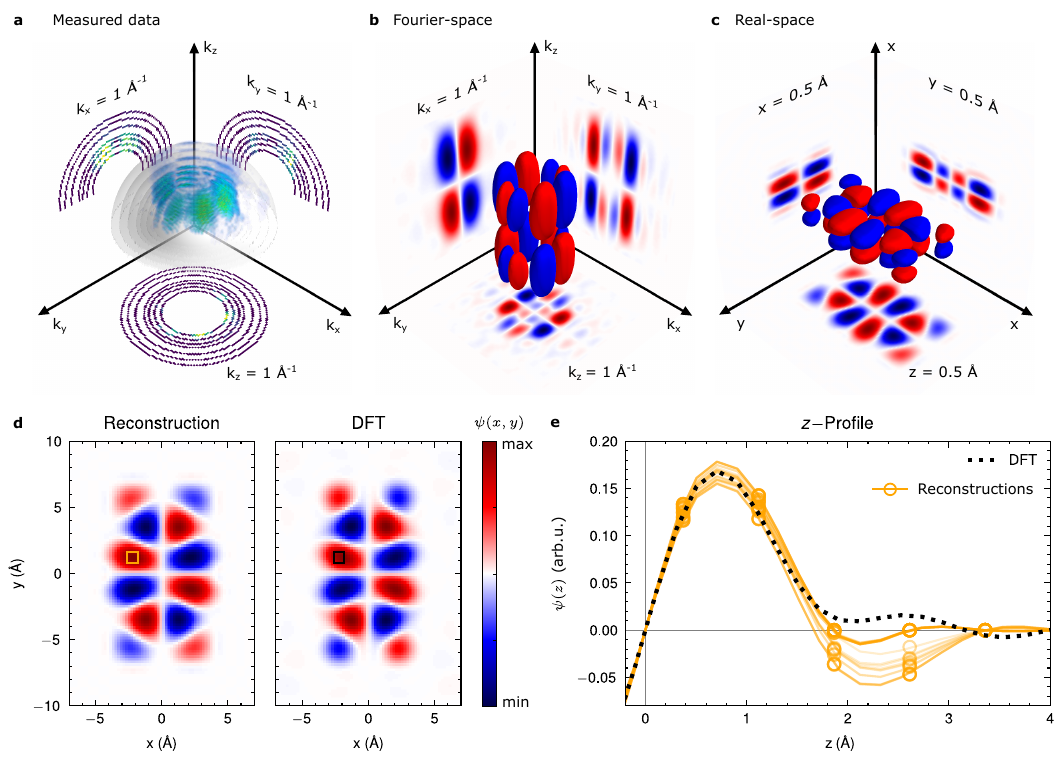}  
    \caption{
    \textbf{3D orbital imaging of the PTCDA highest occupied molecular orbital (HOMO) using only seven photon energies}.
    (a) Here, the recorded momentum-space data is visualized on semi-transparent hemispherical shells (see Methods, Table~\ref{tab:Photonenergies} for the exact photon energies). The sparsity of the data is emphasized by slices through the data at $k_{x/y/z}=1$~\AA\textsuperscript{-1}, showing that only a fraction of the voxels in momentum space contain measurement data. 
    (b,c) The reconstructed HOMO in momentum space and real space, respectively. Evidently, the reconstruction algorithm fully recovers the momentum-space amplitudes and phases, and gives a direct view of the full 3D molecular orbital. 
    (d) Comparison of the in-plane ($x,y$) structure of the reconstructed orbital with the prediction from density functional theory (DFT) of the gas-phase molecule. For an equal comparison, a low-pass filter was applied to the orbital matching to the highest accessible momentum in the experiment.
    (e) Comparison of the reconstructed $z$-dependence of the HOMO (yellow) with DFT (black dashes), extracted at $(x,y)=(-2,1)$~\AA{} (see d). Since each random initialization of the reconstruction algorithm yields a slightly different orbital, we plot here the $z$-dependence of 30 independent reconstructions with transparency set to 20\% (more saturated yellow thus corresponds to multiple reconstructions yielding the same value). Overall, we observe an excellent agreement between experiment and DFT in both the in-plane structure and the $z$-dependence, with particularly good agreement concerning the position of the maximum at $z=0.8$~\AA{}.
    }
    \label{fig:reconstruction_full}
\end{figure}

\section{3D reconstruction of the orbitals}

In order to carry out efficient and time-resolved 3D-POT, the development of a tailored reconstruction algorithm for the 3D orbital from the hemispherical shell photoelectron data is indispensable. Indeed, the most generally applicable method that has been demonstrated to date \cite{graus_three-dimensional_2019} relies on dense photon energy sampling to acquire POT data on hemispherical shells that are spaced closely enough that linear interpolation may be used. However, as argued in Ref.~\cite{dinh_minimalist_2024}, interpolation of the intensity-only POT data ($\propto|\mathcal{F}(\psi)|$) is susceptible to systematic errors around zero crossings in the 3D momentum space distribution $\mathcal{F}(\psi)$. 
Such errors can be completely avoided if the reconstruction of $\mathcal{F}(\psi)$ is incorporated in an algorithm that recovers both amplitude and phase (i.e., sign) simultaneously. 
It is also straightforward to expect that such an algorithm for 3D orbital reconstruction might require less data overall to achieve similar results, which has a direct impact on the experimental design and measurement duration. Indeed, we have recently reported such a minimalist-approach algorithm, which was tested on simulated data~\cite{dinh_minimalist_2024} and we have here further developed the algorithm to be applied to our first experimental 3D-POT momentum microscopy data set.

To recover the full 3D momentum space distribution, including phases and amplitudes for those voxels that were not measured directly (cf. Fig.~\ref{fig:reconstruction_full}), we use the cyclic projections (CP) algorithm that was presented in Ref.~\cite{dinh_minimalist_2024}. In addition to the experimental data, the algorithm incorporates several constraints on the reconstructed orbital. Namely, we employed the known symmetry in the $x$, $y$ and $z$ directions, a loose support of $12 \times 18 \times 6$~\AA\textsuperscript{3} (i.e., roughly matching with the van der Waals size of the molecule) and voxel sparsity of the molecular orbital (800 voxels, i.e., 26\% of the support). The momentum cutoff was set to $4$~\AA\textsuperscript{-1}, corresponding to the photoemission horizon radius for the data measured with the maximum photon energy of 63.8~eV. In Fig.~\ref{fig:reconstruction_full}, we show exemplary reconstruction results of the full 3D molecular orbital for a data set comprising seven photon energies (see Table~\ref{tab:Photonenergies} for details). Fig.~\ref{fig:reconstruction_full}a shows the original sparse 3D-POT data for only seven measured hemispheres, and Fig.~\ref{fig:reconstruction_full}b,c the reconstructed momentum space and real space distributions.
We note that we observe an identically good reconstruction of the LUMO (see Supplementary Fig.~\ref{fig:reconstruction_full_LUMO}).

Having imaged the HOMO with full 3D resolution, we take advantage of the fact that the static non-excited HOMO/LUMO orbitals of PTCDA/Ag(110) can be well approximated using density functional theory (DFT) calculations of the isolated (gas-phase) molecule \cite{luftner_imaging_2014, graus_three-dimensional_2019}, and directly compare our reconstructed orbitals with these theoretical predictions (an identical analysis of the LUMO reconstruction is presented in the Supplementary Information).
In Fig.~\ref{fig:reconstruction_full}d,e, we present a comparison of the reconstructed HOMO orbital with the Kohn-Sham (KS) orbital of the gas-phase molecule calculated by DFT~\cite{puschnig_molecular_database}. It is important to note that the reconstructed orbitals from 3D-POT are intrinsically limited due to the finite momentum range that is accessible in the experiment. At a maximum photoelectron kinetic energy of $63.8$~eV, the intrinsic spatial resolution is approximately $0.75$~\AA{}. For an equal comparison, we therefore apply a low-pass filter matching to this kinetic energy to the KS orbitals. As can be seen in Fig.~\ref{fig:reconstruction_full}d,e, this leads to an excellent agreement for both the in-plane ($x,y$) and the out-of-plane ($z$) distribution. Thus, we present the first realization of a table-top 3D-POT experiment, which is, due to the use of ultrashort pulses from HHG, promising to study dynamical processes with femtosecond time resolution, too.

\begin{figure}[htb]
    \centering
    \includegraphics[width=\textwidth]{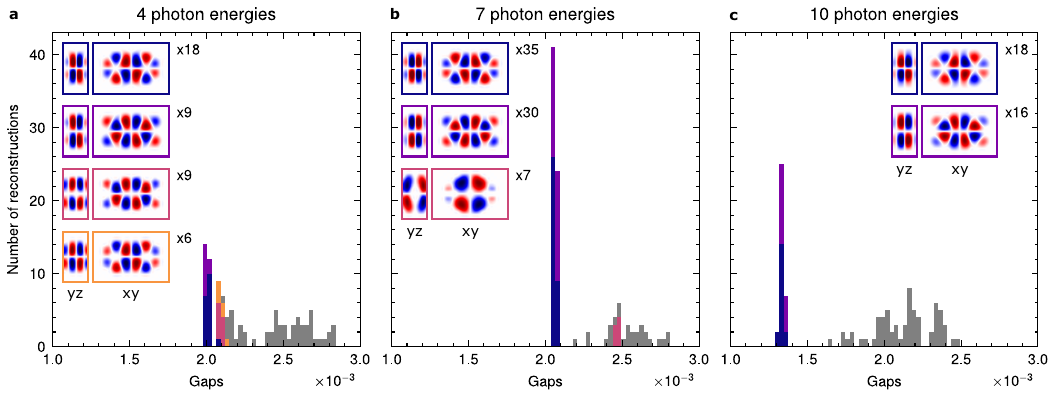}
    \caption{
    \textbf{Comparison of the orbital reconstruction results of the HOMO for 4, 7 and 10 photon energies.} For each data set, the gap of 100 independent reconstructions was analyzed, and DBSCAN clustering (see Methods) was used to determine the most prominent typical recovered orbitals.
    From top to bottom, the insets show vertical (yz) and horizontal (xy) slices through the averaged orbital (i.e., the cluster center) of the retrieved clusters. The colored boxes correspond to the applied support constraint. The gaps of the corresponding reconstructions are colored accordingly in the histogram. Reconstructions which could not be assigned to a cluster are colored gray. 
    a) When only four photon energies are used for the reconstruction, the sparse data leads to an increased number of closely-spaced local minima in the feasibility landscape of the algorithm. Notably, the different reconstructions mostly differ by their z-dependency with varyingly pronounced side peaks (see left insets). 
    b) When seven photon energies are used instead, the global minimum (optimal solution) becomes more defined. Nevertheless, two solutions (blue and purple insets) remain that are only different by a phase flip in momentum space. As discussed in the main text, this is connected to residual background intensity in the momentum maps.
    c) When ten photon energies are used, the stronger measurement constraint leads to an increased gap for faulty reconstructions. The most accurate orbital estimations can therefore easily be selected based upon their small gap value.}
    \label{fig:gaps_and_clusters}
\end{figure}

\section{Accuracy, reliability and experimental design for time-resolved 3D-POT}
Handling of the complex and vast amount of experimental momentum microscopy data is made possible by the application of a powerful algorithm for 3D orbital image reconstruction that not only recovers the unknown phase but also accurately estimates the not-measured amplitudes in momentum space. However, it is known that not every random initialization of the CP algorithm leads to the same local minimum in the algorithm's feasibility landscape (i.e., the estimate of the molecular orbital) \cite{dinh_minimalist_2024}. In fact, it is likely that many local minima exist, and the image constraints and algorithm settings must be considered carefully to ensure an optimal estimation of the molecular orbital.
 
To address this, we run multiple independent reconstructions and analyze their results in terms of the reconstruction gap (a metric indicating the remaining discrepancy between the various reconstruction constraints). Additionally, we use a clustering algorithm (DBSCAN, see Methods) to extract typical reconstructions from the large, multi-dimensional output data. The results of this analysis are visualized in Fig.~\ref{fig:gaps_and_clusters}a for 3D orbital imaging with data from 4, 7 and 10 photon energies. 
We find that the gap, which indicates the remaining discrepancy between the various reconstruction constraints, strongly correlates with the quality of the reconstruction. This is in agreement with Ref.~\cite{dinh_minimalist_2024}, where the smallest gap was shown to be a good measure for the most likely reconstruction. For the experimental data, however, we find that the minimum gap is not a guaranteed indicator: as shown in Fig.~\ref{fig:gaps_and_clusters}, we find two characteristic reconstructions with similarly small gap that differ in the position and shape of the outer lobes (e.g., the blue and purple insets in Fig.~\ref{fig:gaps_and_clusters}). In momentum space, this difference corresponds to a phase flip of the weaker amplitude peaks, which is potentially related to the accuracy of the background subtraction (see Supplementary Fig.~\ref{fig:phaseflip} and \cite{jansen_efficient_2020}). 
Here, we focus on those reconstructions where the momentum-space phase flips between adjacent lobes, since we estimate that the faulty reconstructions result from residual background noise that obscures the true zero amplitudes between these lobes.
Incidentally, this choice is also in line with gas-phase DFT calculations. Taking this into consideration, we emphasize that by analysing the gap and the reconstructed momentum distribution, 3D orbital imaging can be performed without the need to perform additional DFT calculations. 

The amount of local minima in the algorithm's feasibility landscape and the spread in their gap values can be directly influenced by the various constraints that are applied in the algorithm. A straightforward approach is to add experimental data: Going from 4 to 7 or 10 photon energies, this leads to a more well-defined global minimum, reducing thereby the spread in close-to-optimal reconstructions. Moreover, the additional data further penalizes incorrect reconstructions, leading to a larger gap for such reconstructions. This is particularly visible for ten photon energies (Fig.~\ref{fig:gaps_and_clusters}c), where the faulty reconstructions (gray) are spread over a larger range of higher gaps. With more data, it is therefore better possible to discriminate between good and bad reconstructions based upon the gap metric. This data therefore shows that there is a trade-off between reconstruction reliability and experimental effort: more experimental data increase the reliability. For the present static measurements of the PTCDA/Ag(110) HOMO and LUMO orbitals, we find that measuring at seven photon energies sets a good balance between short experiment time and reconstruction reliability.

On the other hand, the feasibility landscape can also be modified by adapting the support, sparsity and symmetry constraints. For example, reduction of the support to a more restrictive shape (e.g., $10.5 \times 16.5 \times 4.5$~\AA\textsuperscript{3}, cf. Supplementary Fig.~\ref{fig:gaps_and_clusters_tightSupport}) strongly reduces the spread in reconstructions for the four photon-energy data set and leads to results which are similar to those for seven photon energies with the van der Waals support. 
Finally, we note that also the choice of optimization algorithm (here CP) will affect the retrieved set of local minima and more advanced algorithms based on the relaxed Douglas-Rachford algorithm may improve convergence to the global minimum \cite{luke_relaxed_2004, LukeDinhJansen2024, LukeDinhJansenMathias2024}. 

\section{Outlook}
The above discussion is most critical and important for the experimental design of a time-resolved 3D-POT experiment, which clearly is possible using our short-pulse HHG-based setup. 
At the femtosecond timescale, where the prediction of excited non-equilibrium orbital wavefunctions is particularly challenging \cite{bennecke_hybrid_2024, caruso_2025roadmap_2025, gonzalez_oliva_hybrid_2022}, the objective is commonly to follow dynamical changes in the electronic structure and orbitals with just a sufficiently amount of data per time-step in order to keep the overall measurement time in a feasible range. In our particular case, we can directly see that after a thorough measurement of the static orbitals with about seven measured momentum hemispheres, this knowledge provides a valuable starting point for the dynamic, i.e. time-resolved, measurements. In other words, while a measurement at seven photon energies is necessary to determine the static orbital and its support, the knowledge extracted from this static characterization, e.g. on the support, can be used to constrain or guide time-resolved 3D-POT with just four photon energies. This obviously leads to a tremendous reduction of the total required data collection and thus measurement time. 
Similarly, for studies of hybridization in, e.g., organic-inorganic heterostructures, an analysis of the available prior knowledge such as the support and symmetry can help to identify the minimum sampling in momentum space and thus the minimum required set of photon energies.

In conclusion, we have demonstrated that full 3D photoemission orbital tomography can be realized efficiently with a table-top ultrashort HHG-based EUV light source and independently of density functional theory calculations. We supplemented this setup by tailored orbital reconstruction algorithms, which allow efficient 3D data extraction and ultrafast time-resolved studies in the future. We have achieved full 3D orbital images of the HOMO and LUMO of PTCDA/Ag(110) and found that this can already be achieved using a data set based on only four photon energies, which we recorded over a total timespan of only 8 hours. With intrinsic femtosecond time resolution, this result positions 3D-POT as a prime candidate, e.g., for the study of light-matter interaction and exciton dynamics in organic semiconductors \cite{neef_orbital-resolved_2022, bennecke_multiorbital_2023} or the study of orbital hybridization in organic semiconductor hybrid interfaces \cite{bennecke_hybrid_2024, krumland_ab_initio_2024, gonzalez_oliva_hybrid_2022}. Finally, improved access to measurements at different photon energies will also help to further develop POT itself, for instance with respect to the plane-wave approximation and final-state effects~\cite{kern_simple_2023}.

\section{Methods}\label{sec:Supplementary.Material}

\subsection{Ultrafast EUV Monochromator}
\label{sec:MoMichromator}
The presented photon energy-dependent photoemission measurements were carried out using the Göttingen in-house photoemission setup~\cite{keunecke_time-resolved_2020} in combination with a newly added tunable EUV beamline based on high-harmonic generation (HHG) and pulse-preserving EUV monochromator. The overall setup is shown in Supplementary Fig.~\ref{fig:momichromator_setup}.

The broadband EUV light is generated by tightly focusing the fundamental of the laser (1030~nm center wavelength, 35~fs pulse duration, 500~kHz repetition rate, 130~\textmu J pulse energy) using a $f = 7.5$~cm lens into an argon jet yielding photon energies $>$71~eV. Directly after the HHG source, a combination of a 4~mm aperture and an anti-reflection-coated EUV/IR separation mirror are used to reject most of the fundamental radiation. The Nb\textsubscript{2}O\textsubscript{5} top coating of the separation mirror allows a high reflection of the EUV radiation, while the dielectric coating ensures an overall IR reflectivity below 0.5\%. The actual monochromator consists of two gold-coated toroidal mirrors ($f=30$~cm, 5\textdegree{} grazing incidence) and a plane grating (5\textdegree{} grazing incidence). The first toroidal mirror collimates the EUV beam, while the second focuses the grating output to a wavelength selection slit. A combined translation/rotation stage allows us to select one of four different plane gratings (100, 300, 400, 600 lines/mm, respectively, see Table~\ref{tab:grating_specs}). 
The rotation stage is set up such that the center of rotation is on the grating surface, which allows optimal wavelength tuning without the need for realignment. These four different grating configurations allow to choose the effective spectral resolution, and thereby to minimize temporal broadening while still ensuring the selection of only one high harmonic. The calculated acceptance bandwidth of the 100~$\mu$m wavelength selection slit and the expected pulse-front tilt (assuming a worst-case HHG source divergence of 20 mrad) are shown in Supplementary Fig.~\ref{fig:MonoSpecs}.

\begin{table}[ht]
    \centering
    \begin{tabular}{|c|c|c|c|c|}
    \hline
    & AOI($^\circ$) & Line density (lines/mm) & Blaze angle ($^\circ$) & Energy range (eV)   \\ \hline
    Grating 1  & \multirow{4}{*}{5}   &  100 &2.3 &13.2$-$30.10 \\ 
    Grating 2  &   &  300 & 2.8 & 20.1$-$51.8\\ 
    Grating 3  &   &  400 & 2.9 & 47.0$-$59.0\\ 
    Grating 4  &   &  600 & 2.6 & 59.0$-$71.0\\ \hline
    \end{tabular}
    \label{tab:MonoSpecs}
    \caption{Overview of the specifications of the used gratings: Grazing incidence angle, line density, blaze angle, and operating energy range.}
    \label{tab:grating_specs}
\end{table}

After the wavelength-selection slit, we use a single $f=80$~cm (6\textdegree{} grazing incidence) toroidal mirror to image the monochromator output to the sample stage of our time-of-flight momentum microscope~\cite{keunecke_time-resolved_2020}. A 100~nm thin aluminum film is used to reject the remaining fundamental laser light, and a photodiode (AXUV100G, EQ Photonics) on a translation stage is used to measure the EUV flux. 

We note that the first two toroidal mirrors are mounted horizontally and thus orthogonal to the other optical components of the monochromator (cf. Supplementary Fig.~\ref{fig:momichromator_setup}). This orientation significantly enhances the overall transmission for $p$-polarized EUV light, which is typically used in ARPES experiments, and leads to a rather polarization-insensitive transmission of the monochromator. See Supplementary Fig.~\ref{fig:SI:efficiency} for the predicted efficiency over the full EUV spectrum. In this way, optimal conditions are established for multidimensional ARPES using 25 photon energies spanning from 13~eV to 71~eV.

\subsection{Photon energy-dependent momentum microscopy}
\label{sec:SI:monochromator}
The PTCDA/Ag(110) sample was prepared in the same way as in Ref.~\cite{jansen_efficient_2020}, by sublimation from a Knudsen cell. The Ag(110) substrate was cleaned prior to the sublimation by Ar ion sputtering and thermal annealing, and was kept at room temperature for the sample growth. Low-energy electron diffraction was used to verify the brick-wall PTCDA superstructure. 

The presented data were acquired using ten different photon energies between 20.5 and 63.8~eV. At each photon energy, we measured a three-dimensional data set depending on the kinetic energy and both parallel momenta of the photoemitted electrons (cf. Ref.~\cite{keunecke_time-resolved_2020}). Before further analysis, the data were calibrated in energy and momentum and corrected for distortion. 
To this end, the data for different photon energies were first aligned in momentum based on their cross-correlation. 
Second, the momentum calibration and distortion were determined by fitting the photoemission horizon of the data set measured with $h\nu=20.5$~eV with an ellipse and subsequently mapping this ellipse on a circle with a fixed radius $0.514\sqrt{E_{\rm kin}}$ for each data slice with kinetic energy $E_{\rm kin}$. 
Observing that the distortion did not noticeably change over the full measurement duration (i.e., between photon energies), the same parameters were used to calibrate and correct all acquired data sets. Finally, the data were intensity-calibrated based on the photon flux measured with the EUV diode and the total accumulation time. 

To determine the energy resolution of each photoemission data set, the spectrum was fitted by the Fermi-Dirac distribution accounting for the thermal broadening at $T=300$~K. The momentum resolution was determined by fitting an error function at the photoemission horizon of the momentum map of the HOMO. For photon energies higher than 50~eV the photoemission horizon was no longer measured due to clipping inside the momentum microscope. The relevant parameters for each measured photon energy (i.e., used grating, energy- and momentum resolution of the photoemission data set, intensity calibration factor) are summarized in Table~\ref{tab:Photonenergies}. Furthermore, the used photon energies for the reconstructions presented in Figs.~\ref{fig:reconstruction_full} (7~$h\nu$) and \ref{fig:gaps_and_clusters} are also indicated.

\begin{table}[ht]
    \centering
    \begin{tabular}{|c|c|c|c|c|c|c|c|}
    \hline
    \parbox{2.cm}{\centering Photon \\energy \\ (eV)} & \parbox{2cm}{\centering Grating\\(lines/mm)} & \parbox{2.cm}{\centering Energy \\ resolution \\ (meV)} & \parbox{2.cm}{\centering Momentum \\ resolution \\ (\AA$^{-1}$) }& \parbox{2cm}{\centering Calibration\\factor} & 4 $h\nu$  & 7 $h\nu$ & 10 $h\nu$ \\ \hline
    20.5 & 300 & 81(3) & 0.073(1) & 1.00(4)     & x & x & x \\
    25.3 & 300 & 58(2) & 0.073(7)& 0.68(3)      &   & x & x \\
    30.1 & 300 & 88(2) & 0.062(3)& 0.90(3)      & x & x & x \\
    34.9 & 300 & 106(2) & 0.077(9)& 1.09(6)     &   & x & x \\
    39.7 & 300 & 129(3) & 0.102(9)& 1.7(2)      &   &   & x \\
    44.5 & 300 & 132(2) & 0.07(2)& 2.8(2)       &   & x & x \\
    49.4 & 400 & 240(5) & 0.13(2)&3.1(1)        & x &   & x \\
    54.2 & 400 & 310(20) & $-$& 6.6(4)          &   & x & x \\
    59.0 & 600 & 202(3) & $-$& 4.6(4)           &   &   & x \\
    63.8 & 600 & 183(3) & $-$& 3.9(4)           & x & x & x \\ \hline
    \end{tabular}
    \caption{\label{tab:Photonenergies} Overview of the parameters of measured photon energies: Used grating, energy and momentum resolution (FWHM) of the respective photoemission data set, calibration, photon energies used for the reconstruction based 4, 7 and 10 photon energies. }
\end{table}

\subsection{Background subtraction}\label{sec:methods:background}
In order to reconstruct a 3D image of the molecular orbitals from the experimental data, it is necessary to extract two-dimensional momentum fingerprints of the individual orbitals and to subtract background contributions, which can for example be due to the $sp$ bands of the silver substrate or due to (in)elastically scattered photoelectrons. 
Here, we present a simple algorithm to extract the molecular orbital fingerprint from each 3D ARPES measurement. The algorithm relies on a simple, but sufficient, modeling of the ARPES spectrum by a Gaussian lineshape of the molecular orbital fingerprint and a background contribution that varies only linearly with the photoelectron kinetic energy. In addition to the description of the algorithm, a Python implementation of the algorithm will also be published along with this paper.

In a first step, we only consider the momentum-integrated photoelectron spectrum $M(E_b)$. In the binding energy ($E_b$) range from 2.5 to 0~eV, the PTCDA/Ag(110) photoelectron spectrum can be modeled as two (quasi-)Gaussian lines superimposed on a linearly-varying background due to the Ag $sp$ bands, modulated by a Fermi-Dirac distribution. Thus, 
\begin{equation}\label{eq:approx.bgr1}
    M(E_b) \approx \sum_{O} \frac{1}{\sigma_O\sqrt{2\pi}}e^{-\frac{1}{2}\frac{(E_b-\mu_O)^2}{\sigma_O^2}}A_O + \frac{A_B + m_B E_{b}}{1+e^{\frac{E_{b}-\mu_B}{\sigma_B}}},
\end{equation}
where the summation over $O$ considers both the HOMO and the LUMO, which is occupied due to charge transfer from the Ag surface. The binding energy, line width, and amplitude are given by $\mu_O$, $\sigma_O$ and $A_O$, respectively. The background is described by a constant offset $A_B$, a slope $m_B$ and the Fermi-Dirac distribution width and position $\sigma_B$ and $\mu_B$. For each photon energy, we determine the parameters $\mu_O$, $\sigma_O$, $\sigma_B$ and $\mu_B$ using the \verb|lmfit| package for Python. These parameters are then fixed for the subsequent momentum-resolved analysis. Note that we observe no photon energy dependence of these parameters, but rather perform the fit for each photon energy to eliminate variations due to differing energy calibrations.

A detailed derivation of the momentum-resolved background subtraction is given in the Supplementary Information, section~\ref{sec:SI:background_algorithm}. Typical retrieved momentum-resolved $A_O$, $A_B$ and $m_B$ are shown in Supplementary Fig.~\ref{fig:bg_subtract}.

From the full results, we conclude that the linear model of Eq.~\eqref{eq:approx.bgr1} enables a sufficient subtraction of the background for 3D orbital tomography. Ultimately, we find that this procedure leaves a small homogeneous background, which we attribute to elastically scattered photoelectrons and disorder in the system. We suppress this background using a simple thresholding operation. An exemplary script to perform this analysis will be published along with this article.

\subsection{3D orbital reconstruction}
\label{sec:SI:algorithm}
In order to reconstruct the full 3D molecular orbital from photoemission data recorded at a limited number of photon energies, it is necessary to recover both the unknown phase in the full 3D momentum space and the unknown amplitudes at momenta ($k$) in between the measured hemispheres. Therefore, we build upon our earlier results \cite{dinh_minimalist_2024}, and use a single iterative reconstruction algorithm to retrieve both at the same time. Similar to Ref.~\cite{dinh_minimalist_2024}, we combine the measured data with prior knowledge in the form of a loose support constraint, a symmetry constraint according to the expected symmetry from density functional theory, and sparsity in the voxel basis. Here, the latter enforces that only a fraction of the voxels within the support have nonzero amplitude.  

For the PTCDA HOMO, we use a box-shaped support of 16 by 24 by 8 voxels, antisymmetry along the $x$, $y$ and $z$ axes, and a sparsity of 800. For the LUMO, we adapt the symmetry constraint to be symmetric along the $y$-axis (the long side of the molecule), but use otherwise identical constraints. 

\subsection{Clustering of the reconstruction results}\label{sec:methods:clustering}
To evaluate the reliability of the reconstruction algorithm, we performed multiple reconstructions for the same parameter set starting from 100 random initializations. To acquire an overview of the resulting data and specifically to extract typical, commonly occurring reconstructed orbitals, we apply a clustering algorithm based on a principal component analysis (PCA) and the DBSCAN algorithm. In a first step, we transform the data using PCA, keeping all components so that no information is lost. Subsequently, we apply DBSCAN density-based spatial clustering (as implemented in the \verb|scikit-learn| package for Python), where the minimum cluster size was set to 6 and the maximal distance between neighboring samples within one cluster was set to 0.012 (0.018) for the HOMO (LUMO). The script used for this analysis will be made available alongside the manuscript.

\clearpage
\section*{Data availability}
The experimental data and reconstructed orbitals will be made available on the GRO.Data platform \cite{benncke_data3D}. 

\section*{Code availability}
Both the code to extract orbital momentum maps from POT data at a single photon energy and the code to reconstruct 3D orbitals from the 3D-POT data will be made publicly available under a Creative Commons BY 4.0 license. 

\section*{Acknowledgments}
This research was funded by the Deutsche Forschungsgemeinschaft (DFG, German Research Foundation) - 432680300/SFB 1456 (project B01), 217133147/SFB 1073 (projects B07 and B10), and 535247173/SPP2244. 

The CNR-IFN activities have been supported by the Italian Ministry of Research (MUR) in the framework of the National Recovery and Resilience Plan “I-PHOQS” Grant (B53C22001750006).

\section*{Author contributions}
M.R., D.St., D.R.L., S.M. and G.S.M.J. conceived the research. F.F. and L.P. designed the EUV monochromator with input from W.B., S.M. and G.S.M.J.. W.B., J.P.B., D.Sch., M.M., L.W., B.v.W., F.F. and L.P. built and aligned the monochromator. W.B. fabricated the sample and performed the measurements. W.B., T.L.D. and G.S.M.J. developed the background subtraction algorithm and T.L.D., D.R.L. and G.S.M.J. developed the reconstruction algorithm. W.B. and T.L.D. performed the 3D reconstructions. All authors discussed the results. D.R.L., S.M. and G.S.M.J. were responsible for the overall project direction. W.B., S.M. and G.S.M.J. wrote the manuscript with contributions from all co-authors. 

\section*{Disclosure statement}
No potential conflict of interest was reported by the author(s).

\clearpage
\renewcommand\thefigure{S\arabic{figure}}
\setcounter{figure}{0} 
\renewcommand{\theHfigure}{Supplement.\thefigure}
\setcounter{equation}{0} 
\renewcommand{\theHequation}{Supplement.\theequation}
\newcommand{\mysection}{\subsection}
\section{Supplementary Information}
The supplementary information includes a schematic figure of the extreme ultraviolet monochromator and figures on the energy resolution, temporal broadening and overall transmission of the monochromator. Concerning the background subtraction routine, a description of the algorithm and a figure with exemplary background subtraction results are given. Furthermore, additional details on the orbital reconstruction are given, including reconstructions with a tighter support and reconstruction of the formerly lowest unoccupied molecular orbital (LUMO) is given. Finally, a discussion of the momentum-space fingerprint of the two most common HOMO and LUMO reconstructions is given.
\mysection{Experimental setup}

\begin{figure}[htb]
    \centering
   \includegraphics[width=\textwidth]{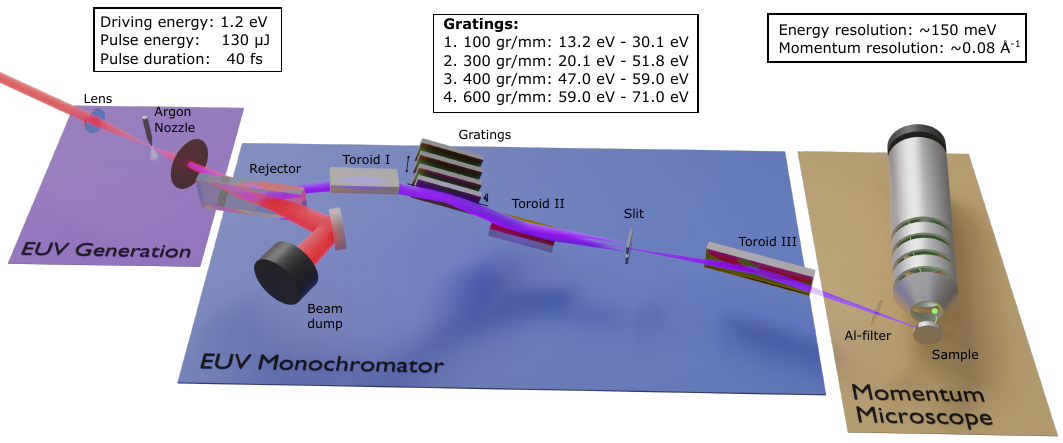}
    \caption{\textbf{Schematic overview of the ultrafast photon-energy-dependent momentum microscopy setup.} On the left, coherent broadband EUV light is generated using 1.2~eV laser pulses focused with $f=7.5$~cm focal length into an argon gas jet. At an anti-reflection-coated mirror for IR, the EUV light is reflected into the monochromator. This consists of 2 toroidal mirrors (operating at 5\textdegree{} grazing incidence) and a set of interchangeable and rotatable gratings mounted in the off-plane diffraction geometry. For each desired photon energy, the least dispersive grating that fully isolates the selected harmonic is chosen to limit the effect of spatial chirp on the pulse duration. The toroidal mirrors I and II collimate and refocus the monochromatized light into a wavelength-selection slit, while a third, long-distance toroidal mirror (III) refocuses the transmitted EUV light to the photoemission momentum microscope. Finally, an EUV-sensitive photodiode can be moved into the beamline just before the last toroidal mirror to measure the monochromatized EUV flux. Specific details on the gratings are given in Methods, section~8.1. } 
    \label{fig:momichromator_setup}
\end{figure}

\begin{figure}[htb]
    \centering
    \includegraphics[width=\linewidth]{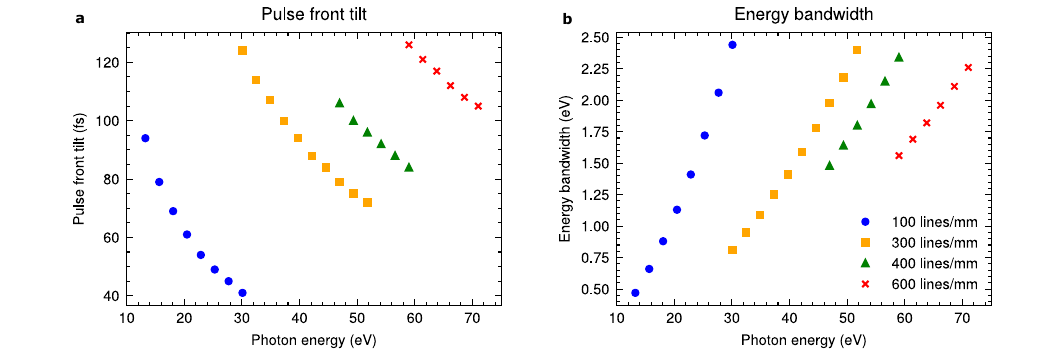}  
    \caption{Temporal broadening due to the pulse front tilt and energy acceptance bandwidth $\rm{\Delta E}$ at the 100~$\mu$m slit. The individual high harmonics can be separated when $\rm{\Delta E} < 2.4$~eV.}
    \label{fig:MonoSpecs}
\end{figure}

\begin{figure}
    \centering
    \includegraphics[width=0.5\linewidth]{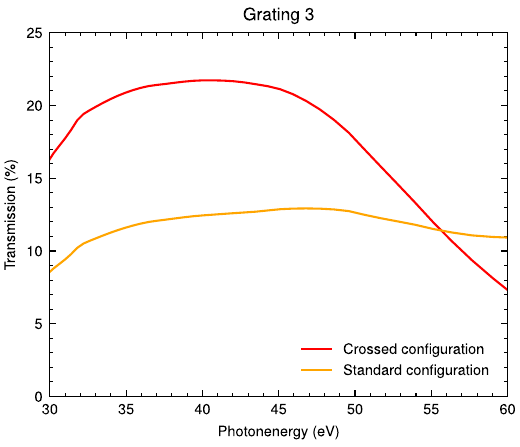}
    \caption{\textbf{Calculated overall transmission of the EUV monochromator.} For simplicity, the initial grazing incidence plate was not included in the calculation, and the grating line spacing was set to 400~lines/mm. The chosen `crossed' configuration, i.e. with two toroidal mirrors mounted horizontally, results in an overall higher transmission for p-polarized light compared to the standard configuration (all optics mounted vertically). Additionally, the `crossed' configuration reduces the polarization dependence of the transmission. The efficiency was calculated using the optical constants of gold, while the grating efficiency was calculated using the Gsolver software. Surface roughness was ignored in the calculation. }
    \label{fig:SI:efficiency}
\end{figure}

\clearpage
\mysection{Momentum-resolved background subtraction algorithm}\label{sec:SI:background_algorithm}
The momentum-integrated analysis of the photoelectron spectrum is described in the Methods, section~7.3. A typical result is shown in Supplementary Fig.~S4a.

For the momentum-resolved analysis, we now consider that $A_O$ and $A_B$ are each momentum-dependent intensity distributions, i.e., $A_O, A_B \in  \mathbb{R}^{N_x\times N_y}_+$, where $N_x$ and $N_y$ are the number of voxels along the $k_x$ and $k_y$ axes, respectively. Also $m_B$ is momentum dependent, but takes both positive and negative values. Since this simple description of the background cannot account for the strongly dispersive band structure of the $sp$ bands observed in this energy region, we constrain the analysis to a small energy region centered on the molecular fingerprint of interest. To extract the HOMO fingerprint, we therefore consider only binding energies between approximately 2.5 and 1.4~eV, while for the LUMO we will consider energies between 1.4 and 0~eV.

Let $M = (M_1,\dots,M_{N_z})$ be a vector where each entry $M_i \in \mathbb{R}^{N_x\times N_y}$ represents the momentum distribution recorded at the i\textsuperscript{th} binding energy. $N_z$ indicates the number of binding energies. Similarly, let $E=(E_1,\dots,E_{N_z})$ indicate the corresponding (real-valued) energies. 
We then rewrite (2) as:
\begin{equation}\label{eq:approx.brg}
    M_i \approx \eta_O^{(i)}A_O + \eta_B^{(i)}A_B + \zeta_B^{(i)}m_B,\, \ \forall i=1,\dots,N_z,
\end{equation}
where $A_O, A_B \in \mathbb{R}^{N_x\times N_y}_+$, $m_B \in \mathbb{R}^{N_x\times N_y}$, and
\begin{equation}
    \eta_O^{(i)}=\frac{1}{\sigma_O\sqrt{2\pi}}e^{-\frac{1}{2}\frac{(E_i-\mu_O)^2}{\sigma_O^2}}\,,\quad 
    \eta_B^{(i)}=(1+e^{\frac{E_{i}-\mu_B}{\sigma_B}})^{-1}\,,\quad \zeta_B^{(i)}=E_i\eta_B^{(i)}.
\end{equation}

We can then define the objective function $g$, acting on a guess of the momentum-dependent distributions $X=(A_O,A_B,m_B)$ as $g(X)=(g_1(X),\dots,g_{N_z}(X))$ with
\begin{equation}
g_i(X)=\eta_O^{(i)}A_O + \eta_B^{(i)}A_B + \zeta_B^{(i)}m_B - M_i,\, \ \forall i=1,\dots,N_z.
\end{equation}
The absolute squared approximation error is then given by $f(X) = \|g(X)\|^2=\sum_{i=1}^{N_z}\|g_i(X)\|^2$. To extract the momentum fingerprints from the data, we therefore consider the minimization problem
\begin{linenomath*}
    \begin{alignat}{1}
        \label{eq:opt.bgr.prob}
        \min_{X\in\Omega}          \quad f(X),
    \end{alignat}
\end{linenomath*}
where the domain $\Omega$ incorporates the non-negativity constraint on $A_O$, $A_B$ and reality constraint on $m_B$. 
Note that problem \eqref{eq:opt.bgr.prob} is convex since the objective function $f$ and the domain $\Omega$ are convex.
Therefore, \ref{eq:opt.bgr.prob} attains only a single minimum value.

We have used the Projected Gradient method to solve \eqref{eq:opt.bgr.prob}. The algorithm is implemented as follows:
Given an initial point $X^{(0)}\in(\mathbb{R}^{N_x\times N_y})^{3}$ (e.g., $X=\mathbf{0}$), $s>0$ small enough, for $n=1,2,\dots,$ do\\
\begin{equation} 
    X^{(n+1)}=P_\Omega \left(X^{(n)}-s\nabla f(X^{(n)})\right).
\end{equation}
Then, by setting a suitable step size $s$, the sequence $(X^{(n)})_{n=1}^\infty$ converges to an optimizer $X^\star$ of problem \eqref{eq:opt.bgr.prob}, see \cite{iusem2003convergence}.

We now compute $\nabla f$ and $P_\Omega$.
Set $f_i(X)= \|g_i(X)\|^2$ for all $i=1,\dots,N_z$. Then we have $f(X)=\sum_{i=1}^{N_z}f_i(X)$, and therefore $\nabla f(X)=\sum_{i=1}^{N_z}\nabla f_i(X)$ with
\begin{equation}
    \nabla f_i(X)=2\begin{pmatrix}
    \cfrac{\partial g_i}{\partial A_O}(X) \cdot g_i(X)\\
    \cfrac{\partial g_i}{\partial A_B}(X)\cdot g_i(X)\\
    \cfrac{\partial g_i}{\partial m_B}(X)\cdot g_i(X)\\
    \end{pmatrix}=2\begin{pmatrix}
    \eta_O^{(i)}\mathds{1}_{N_x\times N_y}\cdot g_i(X)\\
    \eta_B^{(i)}\mathds{1}_{N_x\times N_y}\cdot g_i(X)\\
    \zeta_B^{(i)}\mathds{1}_{N_x\times N_y}\cdot g_i(X)\\
    \end{pmatrix}\, \ \forall i=1,\dots,N_z.
\end{equation}

\begin{figure}[ht]
    \centering
    \includegraphics[width=\textwidth]{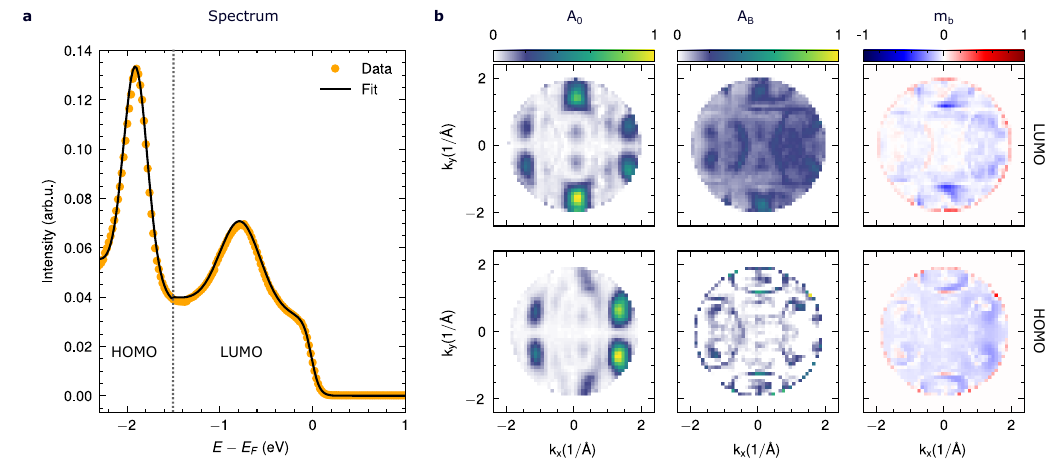}
    \caption{\textbf{Background subtraction for $h\nu=22.5$~eV.} a) Fitted momentum-integrated spectrum according to Eq.~(2) (see Methods). The LUMO and HOMO peak was fitted separately. The corresponding energy ranges are indicated by the dotted gray line. b) Amplitude of the Gaussian peak ($A_0$), the linear background amplitude ($A_B$) and the corresponding slope ($m_b$) for the LUMO (top row) and HOMO (bottom row).}
    \label{fig:bg_subtract}
\end{figure}

Supplementary Fig.~\ref{fig:bg_subtract}b shows typical results retrieved by this procedure.

\begin{figure}[htb]
    \centering
    \includegraphics[width=\textwidth]{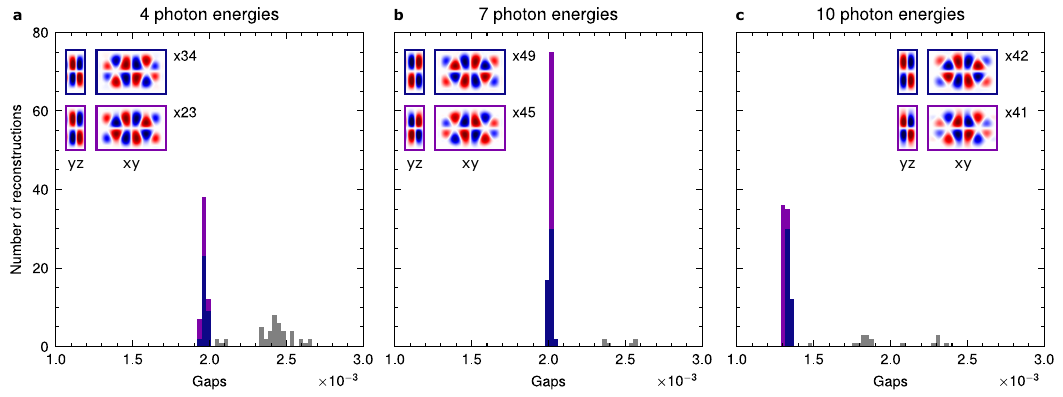}
    \caption{
    \textbf{Comparison of the orbital reconstruction results of the HOMO for 4, 7 and 10 photon energies based on a smaller support size ($10.5 \times 16.5 \times 4.5$~\AA\textsuperscript{3}).} For each data set, the gap of 100 independent reconstructions was analyzed, and DBSCAN clustering (see Methods) was used to determine the most prominent typical recovered orbitals.
    From top to bottom, the insets show vertical (yz) and horizontal (xy) slices through the averaged orbital (i.e., the cluster center) of the retrieved clusters. The colored boxes correspond to the applied support constraint. The gaps of the corresponding reconstructions are colored accordingly in the histogram. Reconstructions which could not be assigned to a cluster are colored gray. 
    In contrast to the van der Waals support, the reconstruction based on four different photon energies (a) leads to a well-defined global minimum with two solutions which differ by a phase flip in momentum space (cf. Supplementary Section \ref{sec:SI:phaseflip}). The same applies for the data sets consisting of seven (b) and ten (c) distinct photon energies, where the number of unassigned reconstructions (gray) is reduced compared to the loose support van der Waals support (cf. Fig. 4).}
    \label{fig:gaps_and_clusters_tightSupport}
\end{figure}

\mysection{Tight support analysis}
The support constraint ($12 \times 18 \times 6$~\AA\textsuperscript{3}) for the orbital reconstruction presented in the main text was chosen to match the van-der-Waals size of the molecule. Here, we discuss reconstruction results for a tighter support constraint ($10.5 \times 16.5 \times 4.5$~\AA\textsuperscript{3}). This increases the reliability of the reconstruction and yields reliable results for the data set consisting of four distinct photon energies. In analogy to Fig.~4, Supplementary Fig.~\ref{fig:gaps_and_clusters_tightSupport} shows the gap distribution and clustering results based on the reconstruction using the tighter support (see Methods section 7.5 for details on the clustering).

\begin{figure}[hp]
    \centering
     \includegraphics[width=\linewidth]{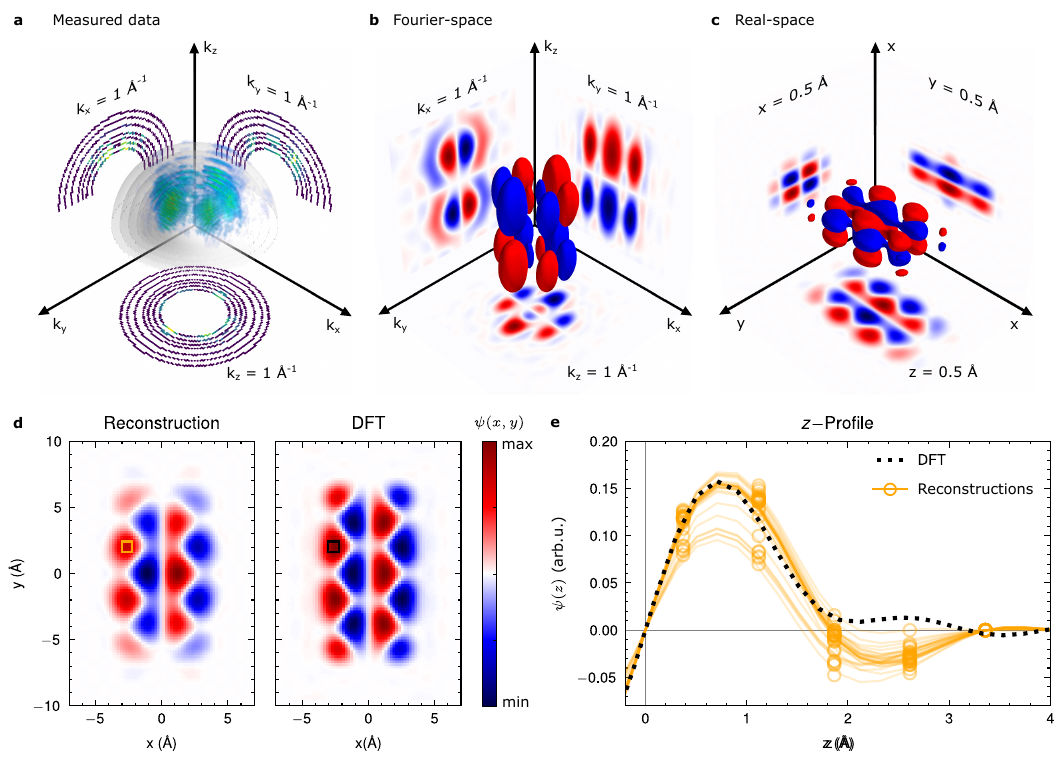}  
    \caption{
    \textbf{3D orbital imaging of the PTCDA lowest unoccupied molecular orbital (LUMO) using only seven photon energies} .
    (a) Here, the recorded momentum-space data is visualized on semi-transparent hemispherical shells (see Methods, Table~2 for the exact photon energies). The sparsity of the data is emphasized by slices through the data at $k_{x/y/z}=1$~\AA\textsuperscript{-1}, showing that only a fraction of the voxels in momentum space contain measurement data. 
    (b,c) The reconstructed LUMO in momentum space and real space, respectively. Evidently, the reconstruction algorithm fully recovers the momentum-space amplitudes and phases, and gives a direct view of the full 3D molecular orbital. 
    (d) Comparison of the in-plane (x,y) structure of the reconstructed orbital with the prediction from density functional theory (DFT) of the gas-phase molecule. For an equal comparison, a low-pass filter was applied to the orbital matching to the highest accessible momentum in the experiment.
    (e) Comparison of the reconstructed $z$-dependence of the LUMO (yellow) with DFT (black dashes), extracted at $(r_x,r_y)=(-2.5,2)$~\AA{} (see d). Since each random initialization of the reconstruction algorithm yields a slightly different orbital, we plot here the $z$-dependence of 30 independent reconstructions with transparency set to 20\% (more saturated yellow thus corresponds to multiple reconstructions yielding the same value). Overall, we observe an excellent agreement between experiment and DFT in both the in-plane structure and the $z$-dependence, with particularly a good agreement concerning the position of the maximum at $z=0.8$~\AA{}.
    }
    \label{fig:reconstruction_full_LUMO}
\end{figure}

\mysection{LUMO analysis}\label{sec:SI:lumo}
Analogous to the reconstruction of the HOMO in the main manuscript, we use the cyclic projection algorithm \cite{dinh_minimalist_2024} with identical parameters to reconstruct the LUMO from the respective momentum maps (cf. Fig.~2). We have run 100 independent reconstructions starting from random initializations using 4, 7 and 10 different photon energies (cf. Table~2). As for the HOMO, the most reliable three-dimensional reconstructions were found to be obtained for seven photon energies. Fig.~\ref{fig:reconstruction_full_LUMO} shows an exemplary reconstruction in momentum- and real-space based on the sparse measured momentum data. The data are compared to the corresponding KS orbital of the gas-phase molecule calculated by DFT (extracted from Ref.~\cite{puschnig_molecular_database}) with the low-pass filter in place. Again, we find excellent agreement between the in-plane ($x,y$) and out-of-plane ($z$) distribution.  

\begin{figure}[hp]
    \centering
    \includegraphics[width=\textwidth]{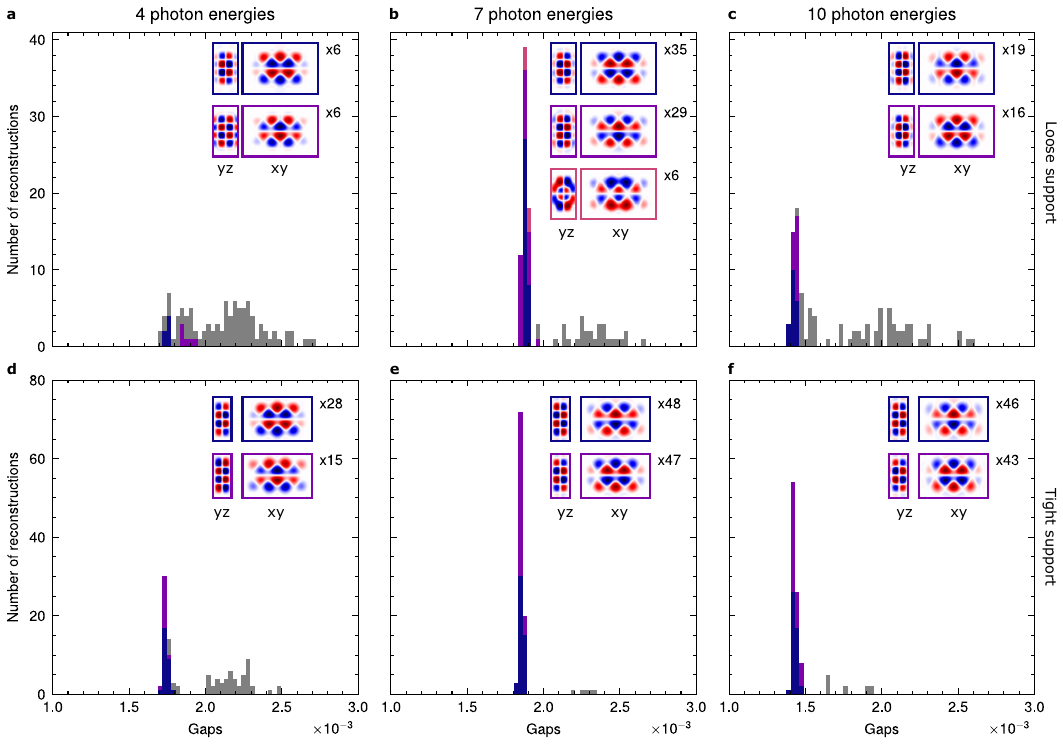}
    \caption{
    \textbf{Comparison of the LUMO reconstruction results for 4, 7 and 10 photon energies based on the loose support (a-c) and the tight support (d-f)}. For each data set, the gap of 100 independent reconstructions was analyzed, and clustering was used to determine the most prominent typical recovered orbitals. Details for the clustering approach are given in the Methods section~8.5. From top to bottom, the insets show vertical (yz) and horizontal (xy) slices through the averaged orbital (i.e., the cluster center) of the retrieved clusters. The colored boxes correspond to the applied support constraint. The gaps of the corresponding reconstructions are colored accordingly in the histogram. Reconstructions which could not be assigned to a cluster are colored gray. The results are comparable to the results obtained for the reconstruction of the HOMO. 
    a,d) When only four photon energies are used for the reconstruction and a loose support ($12 \times 18 \times 6$~\AA\textsuperscript{3}) is applied, the sparse data leads to an increased number of reconstructions which can not be a assigned to a specific cluster and the small gap is not necessarily a good indicator for an accurate reconstruction. The results become more reliable using a tighter support ($10.5 \times 16.5 \times 4.5$~\AA\textsuperscript{3}), where the two most accurate reconstructions (see insets) can be identified by their small gap and only differ by a phase flip in momentum space (cf. Supplementary Section \ref{sec:SI:phaseflip}).
    b,e) When seven photon energies are used instead, the global minimum (optimal solution) becomes more defined for the loose support and the tight support restricts the reconstructions almost exclusively to the two most prominent reconstructions (see insets). 
    c,f) Reconstructions based on ten photon energies exhibit very similar behaviour as for seven photon energies. Again the tighter support increases the reliability of the reconstruction.}
    \label{fig:gaps_and_clusters_LUMO}
\end{figure}

In order to investigate the reliability of the reconstruction based on 4, 7, and 10 distinct photon energies, Fig.~\ref{fig:gaps_and_clusters_LUMO} shows the histogram of the reconstruction gaps. Additionally, the data was clustered using principal component analysis and DBSCAN as described in the Methods section~7.5. Generally, we observe very similar behavior as for the HOMO. Firstly, independent of the used photon energies, we find two most likely reconstructions that can be traced back to a sign change of the phase. Secondly, the gaps consistently increase with the number of photon energies. Thirdly, the reconstruction using seven distinct photon energies yields the most reliable results with the fewest numbers of clusters. These observed trends can be rationalized by similar reasons as those outlined for HOMO in the main text.  

\mysection{Momentum space analysis of the most likely reconstructions}\label{sec:SI:phaseflip}

\begin{figure}[hp]
    \centering
    \includegraphics[width=\textwidth]{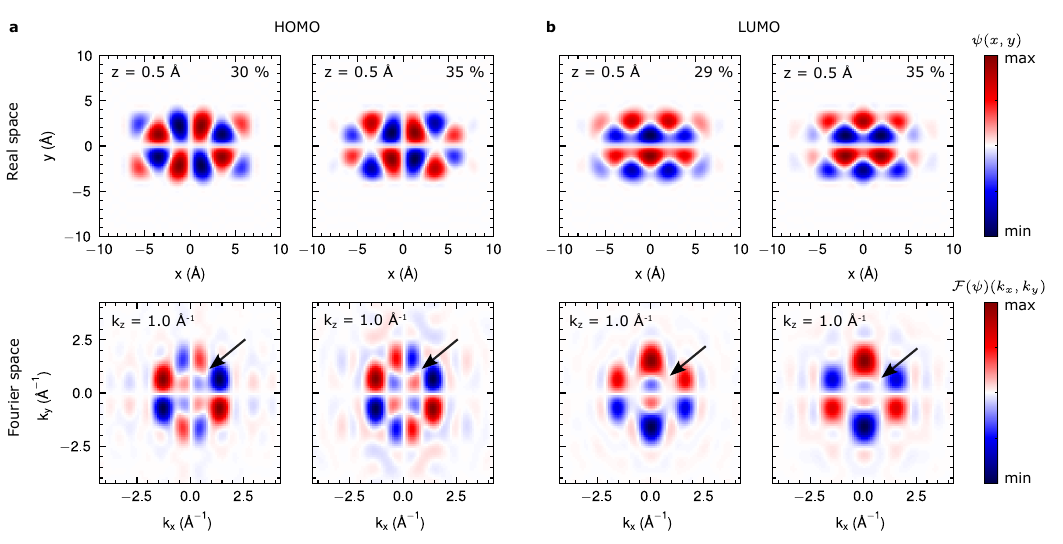}
    \caption{Two most likely reconstructions of the HOMO (a) and the LUMO (b) in real space (top) and momentum space (bottom). The area of the present/absent phase flip in momentum space is indicated by the arrow.}
    \label{fig:phaseflip}
\end{figure}

The employed reconstruction algorithms yields two most likely reconstructions of the HOMO and LUMO (shown in Fig.~4 and Supplementary Fig. \ref{fig:gaps_and_clusters_LUMO}, respectively). Supplementary Fig.~\ref{fig:phaseflip} shows a cut through the real and momentum space representations of the corresponding estimated orbitals. For both the HOMO and the LUMO, the two different reconstruction results differ by a phase flip in momentum space as highlighted by the arrow.

\clearpage

\section*{References}

\bibliographystyle{naturemag}
\bibliography{orbit3d}

\end{document}